# The dynamic adsorption of Xe on a fixed bed adsorber at 77 K

Bin Long [1,2,1)], Jun-Li Li [1], Qun-Shu Wang [2], Shu-Juan Feng [2], Guo-Qing Zhou [2], Tian-Cheng Feng [2], Yan-Jie Tian [2], Huai-Cheng Ma [2]

[1] *Key Laboratory of Particle and Radiation Imaging of Ministry of Education, Department of Engineering Physics, Tsinghua University, Beijing 100084, China*
[2] *Northwest Institute of Nuclear Technology, Xi'an 710024, China*

Abstract: During the design of fixed bed adsorbers, it is vital to understand the dynamic adsorption properties of the system. Because temperature is one of the most important factors affecting adsorbent performance, such that the dynamic adsorption coefficients tend to increase as the temperature decreases, the dynamic adsorption characteristics of Xe on a fixed bed adsorber at 77 K were studied in the present work to minimize the volume of fixed bed adsorber, employing a variety of adsorbents under different operational conditions. The results show that the adsorption performance of carbon molecular sieve is superior to that of activated carbon. And both operational conditions and the presence of gaseous impurities were found to affect adsorption properties.

**Nomenclature**

| | |
|---|---|
| CTBTO | Comprehensive Nuclear-Test-Ban Treaty Organization |
| Xe | Xenon |
| $C_0$ | Equilibrium concentration of inlet gas (V/V) |
| C | Concentration of inlet gas (V/V) |
| LUB | Length of unused bed |
| GC | Gas chromatograph |
| TCD | Thermal conductivity detector |
| $LN_2$ | Liquid nitrogen |
| CMS | Carbon molecular sieve |
| GAC | Granular activated carbon |

## 1. Introduction

Because of their high fission yields, suitable half-lives and low environmental background concentrations, the Xe radionuclides are important with regard to the detection of clandestine nuclear tests as well as in various aspect of radiation safety (C. Zhou et al., 2013). As a result of the typically very low environmental concentrations of these isotopes, they are very difficult to detect unless they are concentrated through separation from other gaseous atmospheric components. The most common method of Xe enrichment through separation is physical adsorption, which has been applied in the CTBTO international monitoring systems and on-site inspection processes, including the ARSA (J. I. McIntyre et al., 2001; T. W. Bowyer et al., 2002; M. Auer et al., 2010), SPALAX (J.-P. Fontaine et al., 2004), ARIX (Y. V. Dubasove et al., 2005), SAUNA (M. Auer et al., 2010; A. Ringbom et al., 2003), SAUNA-OSI, ARIX 3F (V. V. Prelovskii et al., 2007), and XESPM-II techniques (C. Zhou et al., 2013). The adsorption materials used in these systems include carbon-based molecular sieves, granular-activated carbons, and zeolite molecular sieves. Many reports have indicated that adsorption is an exothermal process and the dynamic adsorption capacity of a fixed bed adsorber is inversely proportional to its temperature (C. Zhou et al., 2011; R. T. Yang, 1987). And the dynamic adsorption coefficient of xenon decreased from 2.56L/g to 0.86L/g when the temperature of adsorber increased from 276K to 328K (C. Zhou et al., 2011). Therefore, it is essential to study the properties of Xe on various adsorbents using a cryostat designed to hold the bed temperature constant for minimize the dimension of adsorber of xenon. Liquid nitrogen could be available conveniently, and could refrigerate gases in the adsorber efficiently and rapidly because of its temperature of 77K. In the present work, the dynamic adsorption properties of Xe on a fixed bed adsorber held at 77 K were studied. Consequently, the dynamic adsorption coefficients of different adsorbents and the effects of various operating conditions were assessed.

---

1) Email: longb13@mails.tsinghua.edu.cn

## 2. Theory

The s-shaped breakthrough curve shown in Fig. 1 is one of the most important factors considered during adsorption analysis. This plot is generated from measurements of the concentrations of various gases exiting from the fixed bed adsorber. It presents the relative effluent concentration, $C/C_0$, as a function of either time or specific adsorption volume, expressed as the ratio of the effluent gas volume to the bed capacity. In the case of dynamic adsorption, the breakthrough point is defined by three characteristics: steepness, geometric midpoint, and shape (G. O. Wood, 2002; G. O. Wood, 2002). Based on these criteria, three reference points may be identified: the breakthrough point (5%), midpoint (50%), and saturation point (95%). These correspond respectively to the breakthrough time ($t_{0.05}$), equilibrium time ($t_{0.5}$), and saturation time ($t_{0.95}$). Together with the flow rate, gas concentration, fixed packed bed capacity, and adsorbent mass, the dynamic adsorption coefficients allow adsorbed quantities to be calculated, generating information that is important during the design of a fixed bed adsorber.

According to the literature, the dynamic adsorption coefficients may be expressed by equations (1) to (3) (Y. Nakayama et al., 1994; D. P. Siegworth et al., 1972; R. E. Adams et al., 1959).

$$k_{dB} = \frac{F \cdot t_{0.05}}{m} \quad (1)$$

$$k_d = \frac{F \cdot t_{0.5}}{m} \quad (2)$$

$$k_{dS} = \frac{F \cdot t_{0.95}}{m} \quad (3)$$

Here, $k_{dB}$, $k_d$, $k_{dS}$, $F$, and $m$ are the breakthrough dynamic adsorption coefficient, the dynamic adsorption coefficient, the saturation adsorption coefficient, the outlet gas volumetric flow rate, and the adsorbent mass (g), respectively. The parameter $q$ represents the adsorption quantity, and can be calculated as in equation (4)

$$q = \frac{F \cdot t_{0.5} \cdot C_0}{m} = k_d \cdot C_0 \quad (4)$$

where $C_0$ is the equilibrium concentration of the effluent gas.

The LUB is a factor that is affected by flow rate but not by bed length and is important with regard to optimizing the packed bed length. The LUB value can be approximated by equation (5)

$$LUB = (1 - \frac{t_{0.05}}{t_{0.5}})h \quad (5)$$

where $h$ is the actual length of the packed bed.

## 3. Experimental

(1) Experimental platform

The multifunctional test platform designed for dynamic adsorption experiments is shown in Fig. 2.

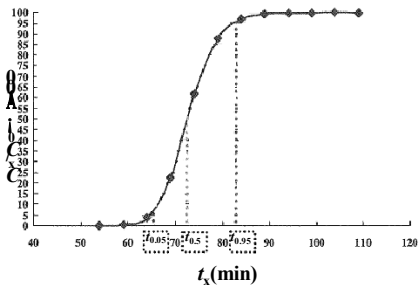

**Fig. 1** A typical breakthrough curve (C. Zhou et al., 2011)

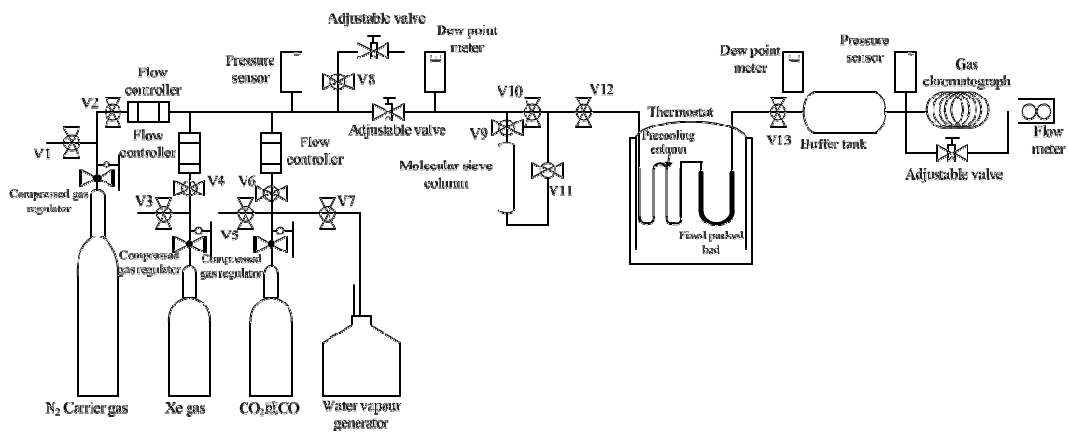

**Fig. 2** Multifunctional test platform for dynamic adsorption experiments

This test platform included a series of high-pressure gas cylinders (containing $N_2$, Xe, $CO_2$, and CO), a water vapor generator, a compressed gas regulator, manual ball valves (V1 to V13), mass flow controllers, pressure sensors, adjustable valves, a molecular sieve column, a thermostat, a pre-cooling column, a packed bed, a buffer tank, dew point meters, a gas chromatograph (GC), and a soap bubble flow meter. $N_2$ and Xe were used as the carrier gas and adsorbate, respectively, while $CO_2$, CO, and water vapor were assessed as gaseous impurities.

The procedure followed during the experimental trials was as follows. Initially, the $N_2$ and Xe gas pressures were set to 0.4 MPa using their respective regulators. The mass flow controllers were adjusted to obtain the desired Xe concentration and gas flow rates so as to ensure that the gases were mixed adequately and homogeneously. The gas pressure was then adjusted using the adjustable valves at the inlet of the fixed bed, after which the Xe concentration of the inlet gas was determined using the GC, employing a 13XHP molecular sieve column to remove water vapor and $CO_2$ before these species could condense inside the low-temperature packed bed. The pre-cooling column and the fixed bed adsorber were placed in a thermostat filled with $LN_2$ at 77 K. Various fixed bed adsorbers were used, made of cylindrical copper tubes having varying dimensions and filled with different adsorbates (hereinafter referred to as the adsorption columns). After the effluent gas from the buffer tank achieved a pressure sufficiently stable to allow for GC analysis, the Xe concentration in the effluent was monitored at 5-min intervals. At the end of the gas circuit, the flow rate was determined with the flow meter and dew point relative humidity was monitored by dew point meters near valves V9 and V13.

The effects of gaseous impurities on Xe adsorption were determined by switching on the pressurized containers of $CO_2$ and CO or the water vapor generator. This was accomplished by closing valves V9 and V11 and opening valve V10.

(2) Adsorbents

The adsorbents included carbon molecular sieves and granular coconut shell-activated carbon. 01- CMS, 501CMS and 601CMS was produced by Shanghai Institute of Fine Chemicals, China. TJ-CMS was manufactured by Tianjin Chemical Reagent Co., Ltd., China. NM- GAC was produced by Zhongsen activated carbon Co., Ltd., Neimeng Province, China. And HN- GAC was procured from Shuangxinlong Industrial and Trading Co., Ltd., Hainan Province, China. The properties of the adsorption columns and adsorbents are summarized in Table 1.

**Table 1** The properties of adsorption columns and adsorbents

| Types | Adsorption column | | Adsorbents | | | | |
|---|---|---|---|---|---|---|---|
| | Size (mm) | Inner diameter (mm) | index | Mass (g) | Mesh (mm) | BET surface area ($m^2/g$) | Pore volume ($cm^3/g$) |
| Carbon molecular sieve | Φ3.175*200 | 1.59 | 01-CMS | 0.2 | 0.25~0.38 | 1049 | 1.04 |
| | Φ3.175*200 | 1.59 | TJ-CMS | 0.26 | 0.25~0.38 | 1129 | 0.50 |
| | Φ3.175*200 | 1.59 | 501CMS | 0.2 | 0.25~0.38 | 1023 | 0.99 |
| | Φ3.175*200 | 1.59 | 601CMS | 0.19 | 0.25~0.38 | 1002 | 0.97 |
| Activated carbon | Φ3.175*200 | 1.59 | NM-GAC | 0.18 | 0.25~0.38 | 1038 | 0.65 |
| | Φ3.175*200 | 1.59 | HN-GAC | 0.22 | 0.25~0.38 | 874 | 0.47 |

(3) Gas chromatography

An HP6890 GC equipped with a TCD was used to analyze the gas concentrations. The operational conditions are provided in Table 2.

**Table 2 GC operational conditions**

| Analysis objects | Oven temperature (℃) | Detector temperature(℃) | Sizes of column | Types of column | Mesh size of column | Column flow (ml/min) | Ref flow (ml/min) | Makeup flow (ml/min) |
|---|---|---|---|---|---|---|---|---|
| $CO_2$, Xe | 80 | 205 | Φ1/8in*2m | Porapak Q | 60~80 | 10 | 26 | 2 |
| CO, Xe | 100 | 205 | Φ1/8in*2m | 5A | 60~80 | 12 | 33 | 2 |

# 4. Results and discussion

(1) Dynamic adsorption of Xe on different adsorbents

The dynamic adsorption characteristics of Xe on the various adsorbents were determined using the gas circuit shown in Fig. 2, with the results presented in Table 3.

**Table 3 The dynamic adsorption characteristics of Xe at 77 K**

| Adsorbents | $C_0$ (×$10^{-6}$V/V) | F (ml/min) | $t_{0.05}$ (min) | $t_{0.5}$ (min) | $k_{dB}$ (L/g) | $k_d$ (L/g) | q (ml/g) | LUB (cm) |
|---|---|---|---|---|---|---|---|---|
| 01-CMS | 39.07 | 515 | 142 | 482 | 365.8 | 1242 | 48.51 | 14.1 |
| TJ-CMS | 39.07 | 491 | 157 | 577 | 385.5 | 1417 | 55.35 | 14.6 |
| 501CMS | 38.64 | 510 | 146 | 551 | 372.2 | 1405 | 54.27 | 14.7 |
| 601CMS | 43.04 | 508 | 135 | 275 | 360.6 | 735 | 31.62 | 10.2 |
| NM-GAC | 38.64 | 547 | 110 | 200 | 301.1 | 547 | 21.15 | 9.0 |
| HN-GAC | 38.64 | 529 | 62 | 247 | 164.1 | 654 | 25.26 | 15.0 |

As can be seen from Table 3, the $k_{dB}$ and $k_d$ values of the carbon molecular sieves are higher than those of the activated carbon. In the case of the carbon molecular sieves, the $k_{dB}$ and $k_d$ values for the TJ-CMS are the highest, at 385.5 and 1417L/g, respectively. Among the activated carbons, the $k_{dB}$ of the NM-GAC is greater than that of the HN-GAC, although its $k_d$ is lower. The LUB values of the 01-CMS, TJ-CMS, and 501CMS are all approximately 14 cm. The $k_d$ of NM- GAC is 547.4 L/g at 77K which is over 200 times than the $k_d$ at 276K (C. Zhou et al., 2011). It is apparent that multilayer adsorption occurs on the carbon molecular sieves and activated carbons at 77 K, resulting in the much higher adsorption capacities of these materials. The $k_{dB}$ values shown here are important because this factor is commonly used to design the first stage of an adsorption column to ensure high recovery of Xe.

(2) The effects of Xe concentration on adsorption performance

Variations in the Xe concentration also affect the dynamic adsorption performance. The dynamic adsorption of Xe on TJ-CMS was thus assessed at varying Xe concentrations, with the results presented in Figs. 3 and 4.

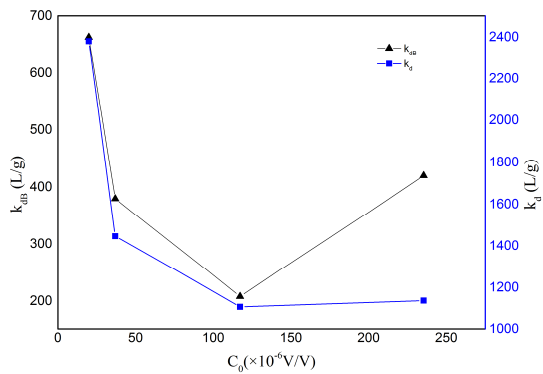

**Fig. 3** The effects of Xe concentration on $k_{dB}$ and $k_d$ over TJ-CMS (flow rate = 492–519 mL/min, Xe concentration 19.9–235×10$^{-6}$ V/V)

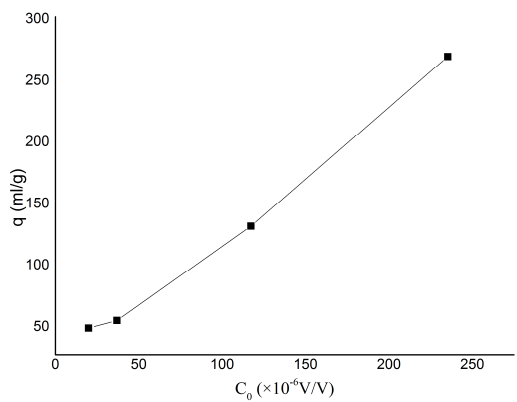

**Fig. 4** Xe adsorption quantity as a function of Xe concentration

As shown in Fig. 3, $k_{dB}$ and $k_d$ initially decrease and then increase as the Xe concentration increases, with the minimum value observed at 117×10$^{-6}$ V/V of Xe. According to Fig. 4, the adsorbed quantity increases as the Xe concentration rises. As the concentration increases, the gases in the adsorption column will tend to liquefy at 77 K and capillary condensation will occur, resulting in the observed adsorption increases.

(3) The effects of flow rate on the dynamic adsorption characteristics of Xe

Using the TJ-CMS carbon molecular sieves, the effects of flow rate on the dynamic adsorption of Xe at 77 K were studied. The test results are presented in Fig. 5.

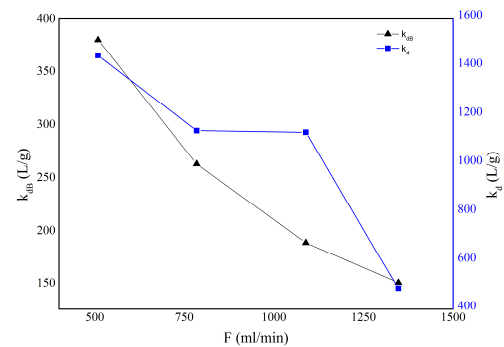

**Fig. 5** Variations in $k_{dB}$ and $k_d$ with the flow rate (flow rate = 509–1348 mL/min, Xe concentration = 37–45 ×10$^{-6}$V/V)

Both $k_{dB}$ and $k_d$ evidently decrease with increasing flow rate. The $k_{dB}$ and $k_d$ decrease from 380 L/g to 150 L/g, from 1500 L/g to 450 L/g, respectively, when the flow rate of gas increases from 509mL/min to 1348mL/min. Because the linear velocity of the gas increases with the flow rate, the gas flowing through the adsorption column may not adequately cool and thus the thermal motion of the gas molecules will increase. Under these conditions, the dynamic adsorption performance will be significantly affected.

(4) The effects of inlet pressure on the dynamic adsorption of Xe

The effects of inlet pressure on the dynamic adsorption of Xe at 77 K were also assessed, with the results shown in Fig. 6.

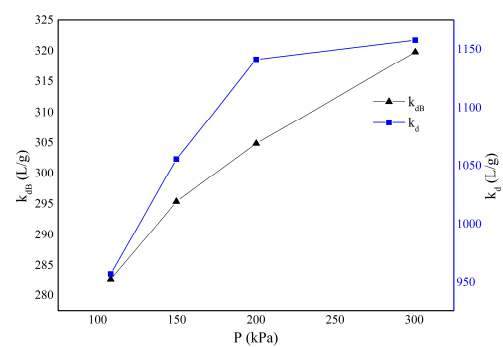

**Fig. 6** The relationships between $k_{dB}$ or $k_d$ and the inlet pressure (P) (flow rate = 565–573 mL/min, Xe concentration = 33–45 ×10$^{-6}$V/V)

Fig. 6 demonstrates that both $k_{dB}$ and $k_d$ increase as the inlet pressure rises, in accordance with previous reports in the literature (R. T. Yang, 1987). Thus, the inlet pressure is one of the most important factors for the design of adsorption columns and gas circuits.

(5) The effects of the inner diameter of the adsorption column on Xe dynamic adsorption

It is well known that the adsorption efficiency of adsorbents will decrease as the linear velocity of the gas increases. For the purpose of reducing the effects of a high linear velocity, the pre-column pressure and the inner diameter of the adsorption column are often increased. Therefore, trials were performed to study the effects of the inner diameter of the adsorption column on Xe dynamic adsorption over the carbon molecular sieve TJ-CMS at 77 K. The resulting data are summarized in Table 4.

Table 4 The effects of adsorption column inner diameter on dynamic adsorption

| Adsorbent mass (g) | Concentration of Xe ($\times 10^{-6}$ V/V) | Flow rate (ml/min) | Linear velocity (cm/s) | Inner diameter of adsorption column (mm) | Pre-column pressure (kPa) | $k_{dB}$ (L/g) |
|---|---|---|---|---|---|---|
| 0.26 | 32.84 | 565 | 476 | 1.6 | 108.2 | 283 |
| 2.1 | 46.97 | 3755 | 452 | 4.2 | 192.7 | 257 |
| 4.18 | 46.97 | 10804 | 496 | 6.8 | 400.7 | 1215 |

According to Table 4, the linear velocities in each trial were approximately equal. Compared with the $k_{dB}$ obtained with a 1.6-mm inner diameter, the $k_{dB}$ for a 4.2-mm inner diameter is nearly the same. However, the $k_{dB}$ increases up to 1215 L/g as the inner diameter increases to 6.8 mm.

(6) The effects of $CO_2$ on the dynamic adsorption of Xe

The effects of different $CO_2$ concentrations on the dynamic adsorption of Xe were examined over the carbon molecular sieve TJ-CMS at 77 K, generating the results shown in Fig. 7. The relationship between inlet pressure and duration when applying 2722 $\times 10^{-6}$ V/V of $CO_2$ is plotted in Fig. 8.

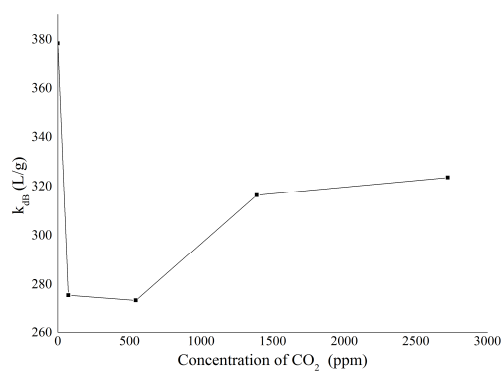

Fig. 7 The relationship between $k_{dB}$ and $CO_2$ concentration (flow rate = 518–576 mL/min, Xe concentration = 37–40 $\times 10^{-6}$ V/V)

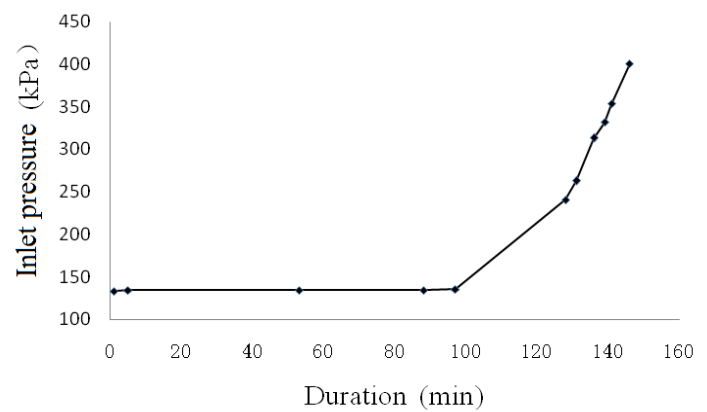

Fig. 8 Pre-column pressure as a function of ventilation duration at 2722 $\times 10^{-6}$ V/V of $CO_2$

As shown in Fig. 7, the $k_{dB}$ value over TJ-CMS decreases and then increases as the concentration of $CO_2$ increases. Because the sublimation temperature of $CO_2$ is 194 K, which is higher than the 77 K bed temperature, the extent of gas liquefaction inside the adsorption column will increase as the $CO_2$ concentration is raised. The inlet pressure will also increase to some degree (Fig. 8). As $k_{dB}$ increases, the flow rate will be reduced owing to condensation of the $CO_2$. Consequently, $CO_2$ must be removed prior to Xe adsorption.

(7) The effects of CO on the dynamic adsorption of Xe

The effects of varying CO concentrations on the dynamic adsorption of Xe were assessed, using the carbon molecular sieve TJ-CMS at 77 K, and the data are summarized in Fig. 9.

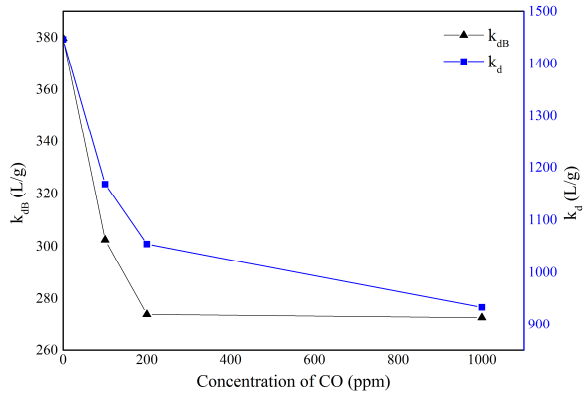

**Fig. 9** The relationship between $k_{dB}$ and CO concentration (Flow rate = 509–554 mL/min, Xe concentration = 35–37 ×10$^{-6}$V/V)

From Fig. 9, it is evident that both $k_{dB}$ and $k_d$ decrease exponentially as the concentration of CO increases, indicating that the presence of CO reduces the adsorption of Xe. Therefore, CO must also be removed before Xe adsorption.

(8) The effects of water vapor on the dynamic adsorption of Xe

According to previous reports (C. Zhou et al., 2013), water vapor tends to reduce the adsorption of Xe, and may also form ice plugs within the adsorption column. The effects of different concentrations of water vapor on the dynamic adsorption of Xe were therefore analyzed, using the carbon molecular sieve TJ-CMS at 77 K. The results are given in Table 5 and Fig. 10. The $k_{dB}$ value over TJ-CMS is seen to decrease as the water vapor concentration increases, thus the presence of water vapor also reduces the adsorption of Xe. Interestingly, the rate of increase in the inlet pressure at 2928 ×10$^{-6}$V/V water vapor is faster than the rate at 553 ×10$^{-6}$V/V. It therefore appears that, at higher concentrations of water vapor, the vapor readily condenses and freezes in the adsorption column, creating a blockage. As such, water vapor must be removed before Xe adsorption.

**Table 5** The effects of water vapor on the dynamic adsorption of Xe

| Moisture content (×10$^{-6}$V/V) | $k_{dB}$ (L/g) |
|---|---|
| 1 | 415.8 |
| 76 | 256.5 |
| 553 | 245.0 |

(Xe concentration=36.6-38.5×10$^{-6}$V/V, flow rate=533-575 mL/min )

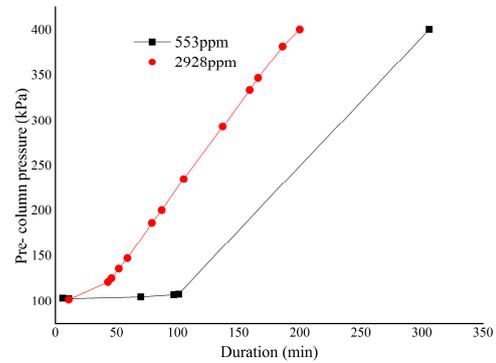

**Fig. 10** Inlet pressure as a function of duration at water vapor concentrations of 553 and 2928 ×10$^{-6}$V/V.

## 5. Conclusions

It is necessary to study the dynamic adsorption properties of a gas prior to designing a fixed bed adsorber. In this paper, the dynamic adsorption characteristics of Xe on various fixed bed adsorbers were studied at 77 K, including the dynamic adsorption coefficients of different adsorbents and the effects of a variety of operational conditions.

The dynamic adsorption performances of carbon molecular sieves are very attractive because they allow us to minimize the adsorption volume at ultra-low temperatures. As such, these materials are superior to activated carbons. The carbon molecular sieve TJ-CMS generated the highest $k_{dB}$ and $k_d$ values among the adsorbents assessed. In this work, the factors affecting dynamic Xe adsorption badly were studied, including Xe concentration, inlet pressure, flow rate, and column inner diameter. Increasing concentrations of $CO_2$, CO, and water vapor reduced the adsorption of Xe because these gases tended to liquefy and even freeze in the adsorption column, blocking the pipeline. Therefore, the removal of gaseous impurities of $CO_2$ and water vapor with zeolite molecular sieve must be considered, because of their higher characteristic energy of adsorption and excellent adsorption capacity for polar molecules. And the operational conditions must be considered during the design of gas circuits.